\documentclass[aps,prl,floatfix,twocolumn,footinbib]{revtex4}
\usepackage{epsfig}
\usepackage[colorlinks=true, citecolor=black, urlcolor=black, linkcolor=black]{hyperref}
\graphicspath{{./}}
\begin{document}

\title{First results of the CERN Resonant WISP Search (CROWS)}
\thanks{Work supported by the Wolfgang-Gentner-Programme of the Bundesministerium f\"ur Bildung und Forschung (BMBF).}
\author{M. Betz}
\author{F. Caspers}
\author{M. Gasior}
\affiliation{CERN Beams Department, CH-1211 Geneva, Switzerland}
\author{M. Thumm}
\affiliation{Karlsruhe Institute of Technology (KIT), IHM and IHE, Kaiserstr. 12, 76131 Karlsruhe}
\author{S. W. Rieger}
\affiliation{University of Geneva, Brain \& Behaviour Laboratory, CH-1211 Geneva, Switzerland}

\begin{abstract}
The CERN Resonant WISP Search (CROWS) probes the existence of Weakly Interacting Sub-eV Particles (WISPs) like axions or hidden sector photons. It is based on the principle of an optical light shining through the wall experiment, adapted to microwaves. 
Critical aspects of the experiment are electromagnetic shielding, design and operation of low loss cavity resonators and the detection of weak sinusoidal microwave signals. 
Lower bounds were set on the coupling constant $g = 4.5 \cdot 10^{-8}$~GeV$^{-1}$ for axion like particles with a mass of $m_a = 7.2~\mu$eV. For hidden sector photons, lower bounds were set for the coupling constant $\chi = 4.1 \cdot 10^{-9}$ at a mass of $m_{\gamma'} = 10.8~\mu$eV. 
For the latter we were probing a previously unexplored region in the parameter space.
\end{abstract}

\maketitle
\section{\label{sec:Intro} Introduction}
Many well motivated extensions to the standard model predict the existence of a new family of particles, the so called Weakly Interacting Sub-eV Particles (WISPs). As the name suggests, they all share a very low rest mass below 1~eV and very feeble interactions with the standard model, making them difficult to detect experimentally.

One popular member of the WISP family is the axion. Historically it emerged from a proposal by Peccei and Quinn in 1977, intended to solve a fine tuning problem in the theoretical framework of Quantum Chromodynamics (QCD) \cite{PhysRevLett.38.1440, PhysRevLett.40.223, Kim19871, PhysRevLett.40.279}. 
Since then, several Axion Like Particles (ALPs) with similar properties to the original axion have been proposed, arising from string theory \cite{1126-6708-2006-06-051}, or motivated as a possible explanation of dark energy in our universe \cite{Raffelt:1996wa}. 
Another prominent member of the WISP family is the Hidden Sector Photon (HSP). It can be described by extra U(1) gauge factors in the standard model, which is a necessary requirement for string theory \cite{Okun:1982xi,src:HSP_STRING,src:HSP_STRING_2,src:LowEnergyFrontier}.

WISPs could explain several astrophysical phenomena \cite{Raffelt:1996wa} and the axion would be an excellent candidate for cold dark matter, if it exists in a certain mass range. Axions also have been the most accepted solution for the strong CP problem in QCD for over 30 years now. However, there is no experimental evidence from laboratory searches yet and all efforts so far have just produced exclusion limits.

\section{\label{sec:exp} Experimental detection principle}
WISPs interact very weakly with standard model particles and their weak coupling to photons provides the most promising way to indirectly observe them in a laboratory experiment.

ALPs can convert to photons and photons can convert to ALPs in a strong static magnetic field by the ``Primakoff effect'' \cite{PhysRevLett.51.1415}. The probability of this process happening is extraordinarily low. 
A conversion by the Primakoff effect happens without energy loss. This means that the entire energy of the photon converts into rest mass ($m_a$) and into kinetic energy of the ALP. The mass of the ALP is a fixed but not yet known parameter, which is only weakly bound by cosmological observations in the range of $10^{-12}~\mathrm{eV} \leq m_a \leq 10^{3} ~\mathrm{eV}$ \cite{Raffelt:1996wa}.
 
The ``Light Shining Through the Wall'' (LSW) detection scheme has first been proposed in \cite{Okun:1982xi,PhysRevLett.59.759,src:hoogLaser}.
These proposals were focused on the design of an experiment in the optical domain. 

A laser beam, shining through a strong magnetic field forms the ``emitting region''. 
In this environment, photons can convert into ALPs, which would propagate parallel to the photon beam. An opaque wall is placed downstream of the magnet, blocking all photons. As the ALP beam does not interact with matter, it penetrates the wall and propagates towards the ``detection region''. A second magnet reconverts the ALPs to photons which can be detected by a sensitive optical instrument. To improve the low conversion probability, mirrors can be placed at each end of the emitting and detection region, forming optical resonators and allowing the photons to pass several times through the magnetic field. This has been done -- for example -- in the ALPS-1 experiment \cite{src:alps1}.

While the coupling between photons and ALPs originates from the Primakoff effect, the coupling between photons and HSPs arises from kinetic mixing, a process similar to neutrino oscillations \cite{src:HSP_STRING}. HSPs can be probed with the same experimental setup as used for ALPs, but -- due to the different coupling mechanism -- a static magnetic field is not necessary for the HSP-search.

\begin{figure}
\centerline{\psfig{file=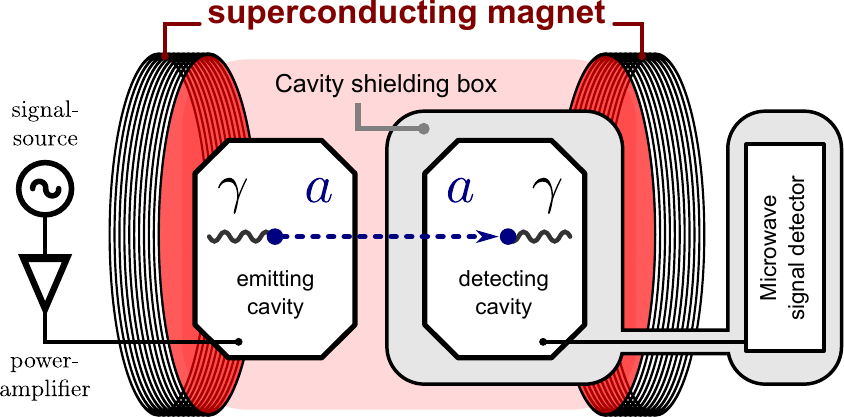, width=1\linewidth}}
\caption{Schematic of a microwave LSW experiment.}
\label{fig:OverViewSimple}
\end{figure}

Adapting the LSW principle to microwaves has first been proposed in \cite{src:hoogUW,src:JaCaRi}.
We performed a microwave based LSW experiment searching for ALPs and HSPs. The schematic of our setup is shown in Fig.~\ref{fig:OverViewSimple}. Converting the optical to a microwave domain setup involves several steps:
The laser is replaced by a microwave oscillator and power amplifier. The equivalent to optical resonators are microwave cavities.
The two dimensional ``wall'' becomes a three dimensional electromagnetic shielding challenge, and 
the optical power detector transforms to a coherent microwave receiver. 

Observing WISPs would correspond to a microwave signal appearing within a well shielded detection volume, exciting the detecting cavity. The weak sinusoidal signal is of equivalent -- and thus of known -- frequency, as the one driving the emitting cavity, allowing us to exploit a lock-in scheme for the signal detection.

Due to the small wavelengths and therefore stringent mechanical tolerances involved, the realization of low loss resonators is a substantial challenge in the optical regime. It is technologically less challenging to produce and operate low loss microwave cavities, making the experimental setup more rugged, cheaper and easier to handle. It also allows to reduce the separation between the cavities to less than a wavelength, making the experiment sensitive -- not only to propagating WISPs (like in a laser based experiment) -- but also to ``near field" WISPs \cite{src:hoogUW} surrounding the cavities. 
The energy required to produce photons decreases as the wavelength increases. For a given input power, more photons are produced in microwave based experiments. Therefore they are more sensitive to WISPs than the optical equivalent. On the downside, the lower photon energy restricts the maximum detectable mass of the hidden particles.

\section{\label{sec:sens}Detection sensitivity}
The expected output power from the detecting cavity due to the ALP or HSP propagation has been derived in \cite{src:JaCaRi} and is given by Eq.~\ref{equ:powerAxions} and Eq.~\ref{equ:powerHSP}:
\begin{eqnarray}
	\label{equ:powerAxions}
	P_{\mathrm{ALP}} &= \left(\frac{g B}{f_{\mathrm{sys}}}\right)^4 |G_{\mathrm{ALP}}|^2 Q_{\mathrm{em}} Q_{\mathrm{det}} P_{\mathrm{em}},
\end{eqnarray}
\begin{eqnarray}
\label{equ:powerHSP}
P_{\mathrm{HSP}} &= \chi^4 \left(\frac{m_{\gamma'}}{f_{\mathrm{sys}}}\right)^8 |G_{\mathrm{HSP}}|^2 Q_{\mathrm{em}} Q_{\mathrm{det}} P_{\mathrm{em}},
\end{eqnarray}
where $Q_{\mathrm{em}}$ and $Q_{\mathrm{det}}$ are the loaded quality factors of emitting and detection cavity, $f_{\mathrm{sys}}$ is the operation frequency of the experiment (to which the cavities are tuned), $P_{\mathrm{em}}$ is the emitting cavity driving power and $B$ is the strength of the static magnetic field. The unknown coupling constants for ALPs or HSPs to photons are given by $g$ and $\chi$ respectively. 
Details of the geometric form factors $|G_{\mathrm{ALP}}|$ and $|G_{\mathrm{HSP}}|$ are given in the subsequent section. 
Both equations are a function of rest mass of the hidden particle, directly by $m_{\gamma'}$ (only in Eq.~\ref{equ:powerHSP}) and indirectly by the mass dependent geometric form factors (in both equations). 
It is convenient to express all quantities in the same unit based on [eV].

If no ALPs or HSPs are observed in the experiment, an upper bound on the coupling parameter $g$ or $\chi$ can be derived from Eq.~\ref{equ:powerAxions} or Eq.~\ref{equ:powerHSP}. This allows to compare the experiments sensitivity to other WISP searches. Note that in this case, the minimum detectable signal power of the RF receiver ($P_{\mathrm{sig}}$) is assigned to $P_{\mathrm{ALP}}$ or $P_{\mathrm{HSP}}$.

Sensitivity to ALPs is largely dominated by the strength of the magnetic field $B$ and the operating frequency $f_{\mathrm{sys}}$. 
Sensitivity to HSPs is dominated by the Q factors of the cavities and the geometric form factor $|G_{\mathrm{HSP}}|$. 

\section{\label{sec:geom}Calculation of the geometric form factor}
The geometric form factor $|G|$ is typically in the order of 1, and depends on the relative position and orientation of the cavities, the electric field configuration of the resonating mode, and the rest mass of the hidden particles. Furthermore it depends on the relative direction of the static magnetic field for ALPs detection \cite{src:JaCaRi, src:JaCaRi2, src:UWA_LSW, src:hoogUW}.
The geometric form factor can be compared to the near-field antenna gain, as used in communication systems. It differs by taking the non-zero rest mass of the WISPs into account.
$|G|$ is determined by a 6 dimensional integration over the volumes of emitting ($V$) and detecting ($V'$) cavity. The formulas are given for ALPs in Eq.~\ref{equ:GALP} and for HSPs in Eq.~\ref{equ:GHSP}:
\begin{eqnarray}
	\label{equ:GALP}
	G_{\mathrm{ALP}} = \frac{k^{2}}{4 \pi} \int_{V'} \int_{V} \frac{e^{i k' \left| \mathbf{x} - \mathbf{y} \right|}}
	{\left| \mathbf{x} - \mathbf{y} \right|} E_{B}(\mathbf{x}) E'_{B'}(\mathbf{y}) ~ d^3 \mathbf{x} d^3 \mathbf{y},
\end{eqnarray}
\begin{eqnarray}
	\label{equ:GHSP}
	G_{\mathrm{HSP}} = \frac{k^{2}}{4 \pi} \int_{V'} \int_{V} \frac{e^{i k' \left| \mathbf{x} - \mathbf{y} \right|}}
	{\left| \mathbf{x} - \mathbf{y} \right|} \mathbf{E(x)} \cdot \mathbf{E'(y)} ~ d^3 \mathbf{x} d^3 \mathbf{y}.
\end{eqnarray}
Each integration variable, $\mathbf{x}$ and $\mathbf{y}$, represents a three dimensional vector, indexing a point within the emitting and receiving cavity in a common coordinate system.
The wavenumber of the photon is given by $k$. The wavenumber of the ALP or HSP is given by $k'$, which depends on the rest mass of the hidden particle ($m_{\mathrm{WISP}}$ in Eq.~\ref{equ:kkk}). This mass dependence has a significant influence on the shape of the excluded areas in Fig.~\ref{fig:exclPlotALP} and Fig.~\ref{fig:exclPlotHSP}.
Both integrands in Eq.~\ref{equ:GALP} and Eq.~\ref{equ:GHSP} contain an attenuation factor proportional to distance ($|\mathbf{x} - \mathbf{y}|$) and a phase factor ($e^{i k' \left| \mathbf{x} - \mathbf{y} \right|}$), which becomes more significant at larger $k'$ (corresponding to WISPs with higher kinetic energy). Note that $k'$ can become complex in the non-propagating WISP case (if $m_{\mathrm{WISP}} > h f_{\mathrm{sys}}$), leading to an exponential suppression of $|G|$.
\begin{eqnarray}
	\label{equ:kkk}
	k = \frac{2 \pi f_{\mathrm{sys}}}{c}	\quad\quad		
	\frac{k'}{k} = \sqrt{1 - \left( \frac{m_{\mathrm{WISP}}}{h f_{\mathrm{sys}}} \right)^2}.
\end{eqnarray}
The $\mathbf{E}$ fields are normalized such that:
\begin{eqnarray}
	\label{equ:normalize}
	\int_{V} \left| \mathbf{E}(\mathbf{x}) \right|^2 d^3\mathbf{x} = 1.
\end{eqnarray}
For HSPs, Eq.~\ref{equ:GHSP} suggests the dot product between the two electric fields, $\mathbf{E(x)} \cdot \mathbf{E'(y)}$, is of importance. It is thus advantageous to use the same mode in both cavities. For ALPs Eq.~\ref{equ:GALP} suggests that only the component of the electric field in the cavity, aligned with the static magnetic field, $E_B$ and $E'_{B'}$ is of significance. Large geometric form factors can be expected, if the electric field is parallel to the external magnetic field over a large volume in the cavity.

\section{\label{sec:cav}Cavity design}
\begin{figure}
\centerline{\psfig{file=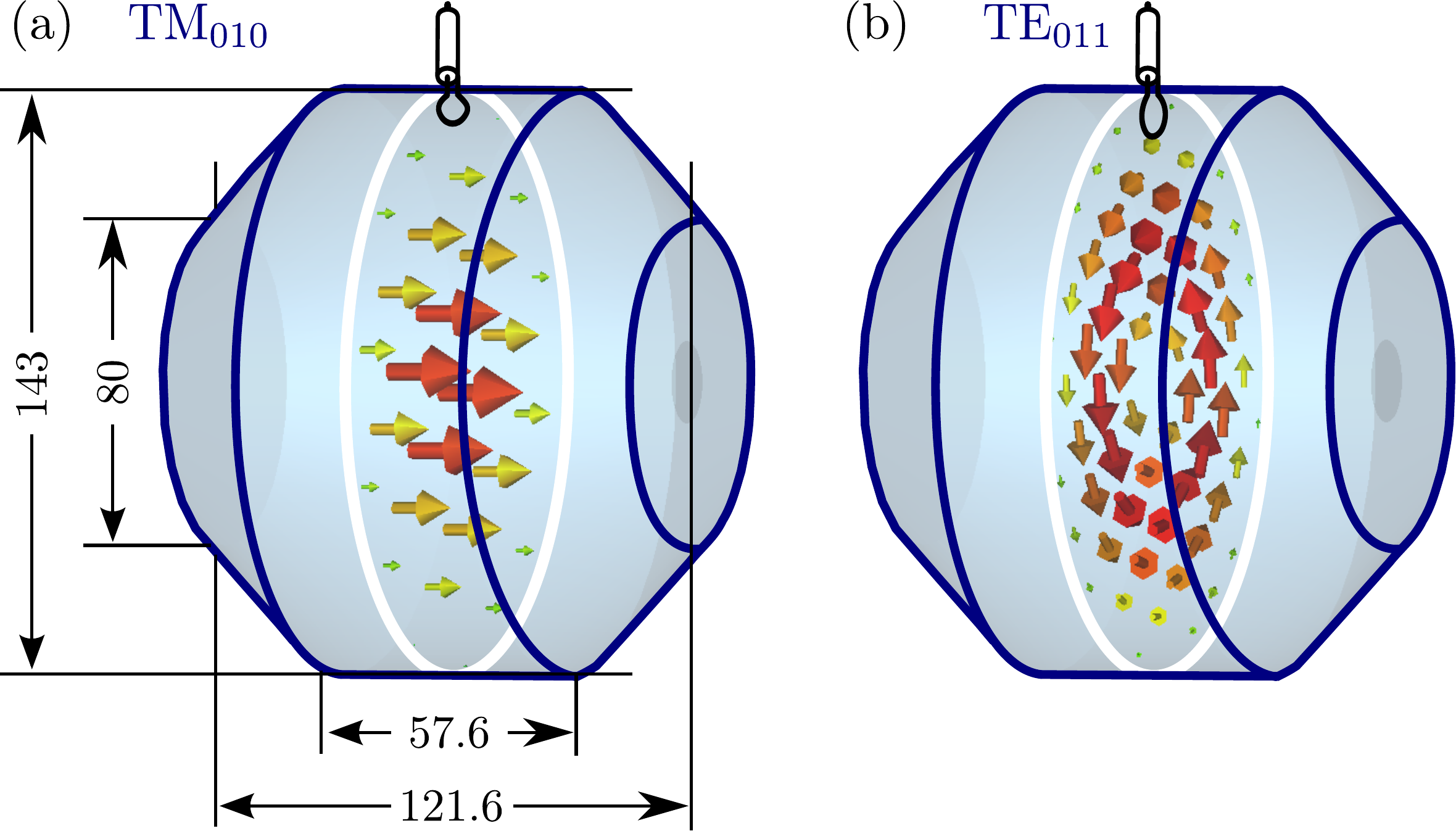, width=1\linewidth}}
\caption{
Inside dimensions of the cavity in [mm]. The coupling loop can be seen on top. The electric field configuration of two modes is shown on a transverse cutting plane. 
(a) TM$_{010}$ mode for ALP search, (b) TE$_{011}$ mode for HSP search.
}
\label{fig:cavSchem}
\end{figure}

Each of the two cylindrical microwave cavities is made out of two half shells, machined at CERN from brass material. Figure~\ref{fig:cavSchem} shows the inner dimensions of the cavity. A photo of the emitting cavity is shown in Fig.~\ref{fig:magnParts}. To increase the surface conductivity, the base material was coated with a 10 $\mu$m thick layer of silver.
On top of the silver layer, a~$\ll 1\mu$m thin flash of gold has been deposited, which serves as protection against oxidation.
Due to the skin effect, $> 80\%$ of the RF currents flow in the low resistivity silver coating.

In order to determine the most sensitive cavity mode for ALP or HSP search, the product of Q factor and corresponding geometry factor needs to be maximized. For ALP search, the best choice is the fundamental TM$_{010}$ mode, providing an E-field which can be aligned with an homogeneous external magnetic field over the largest possible volume, and thus providing a significantly larger $|G|_{\mathrm{ALP}}$ than any other mode.

For the HSP search, the E-field does not need to be aligned with an external magnetic field and as Table~\ref{tbl:g} points out, the TE$_{011}$ mode is preferable. Its displacement currents flow along the circumference of the cavity walls \cite{src:cavTE011lowLoss} and hence do not cross the boundary between the two half shells. Resistive losses due to contact springs are effectively avoided, and therefore its Q factor is higher compared to other modes.

The surface currents of this mode flow entirely azimuthal and do not cross the boundary between the two half shells of the cavity \cite{src:cavTE011lowLoss}. Resistive losses due to contact springs are effectively avoided and therefore its Q factor is higher compared to other modes.

\begin{figure}
\centerline{\psfig{file=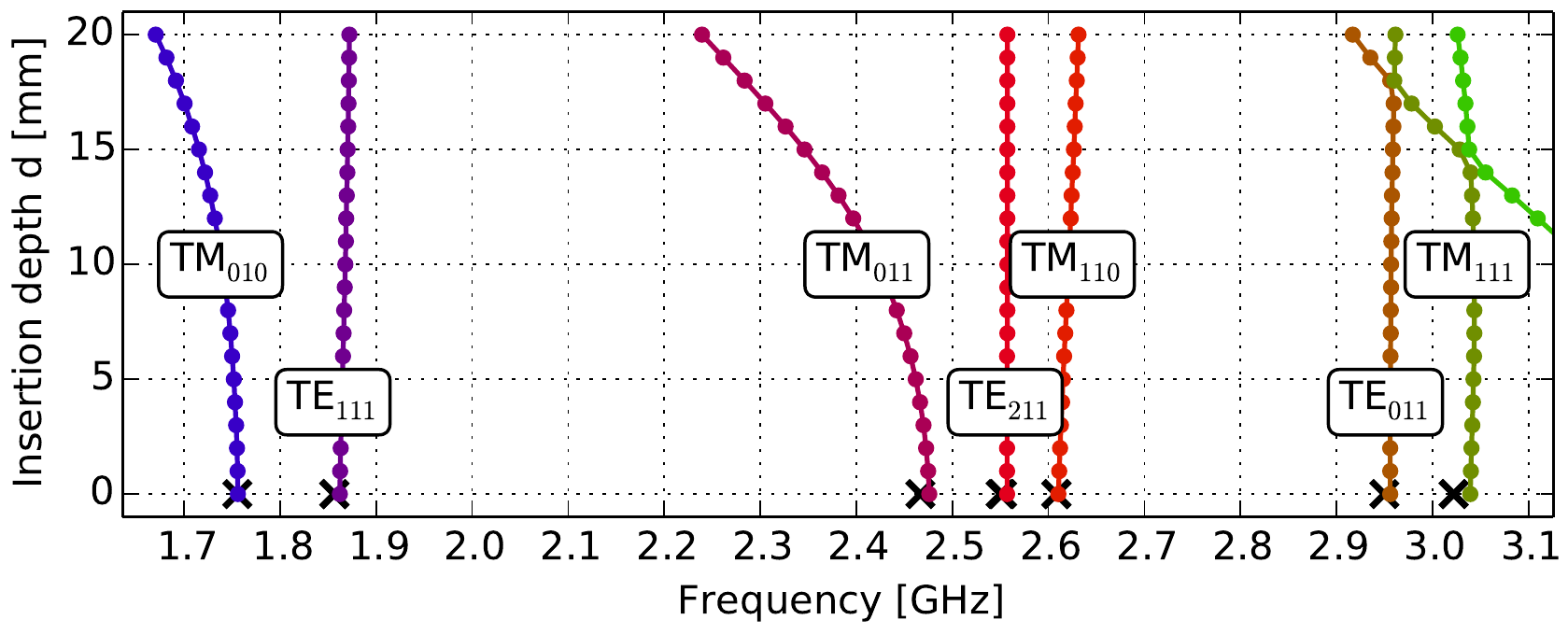, width=1\linewidth}}
\caption{
Tuning range of several modes in the cavity, measured with a VNA in reflection. The depth of the tuning screw $d$ has been increased by 1 mm for each measurement. Black crosses indicate complementary simulation results.
}
\label{fig:cavityTuningMeas} 
\end{figure}

\begin{table}[t]
\centering
\caption{Comparison of modes for HSP search}
\label{tbl:g}
\begin{tabular}{p{0.2\linewidth}p{0.2\linewidth}p{0.2\linewidth}p{0.2\linewidth}}
\hline
Mode & meas. $Q_L$ & $|G|_{\mathrm{HSP}}$ & $Q \cdot G$\\
\hline
TM$_{010}$ & 11 392 & 0.77 & 8772\\
TE$_{011}$ & 23 210 & 0.52 & 12069\\
\hline
\end{tabular}
\end{table}
In a cavity of cylindrical geometry, the TE$_{011}$ and TM$_{111}$ are degenerate and can not be excited separately. To ensure well defined experimental conditions, the geometry has been modified. Chamfering the edges of the cavity breaks the degeneracy between the TE$_{011}$ and TM$_{111}$ mode, separating them in frequency and mitigating energy loss as a consequence of mode coupling \cite{src:cavDegPaper}.
The nominal resonant frequency can be tuned in a range of +- 5 MHz, using a fine threaded tuning screw which modifies the fields in the cavity. A measurement of the frequency dependence of the first seven modes as a function of the tuning screw position is shown in Fig.~\ref{fig:cavityTuningMeas}. The figure also indicates that there are no mode crossings within the nominal tuning range of 0 - 10~mm insertion depth.
A wire-loop antenna of $\approx 8$~mm diameter couples the electromagnetic field in the cavity to a $50~\Omega$ coaxial transmission line. 
The coupling strength ($\beta$) of this loop antenna can be easily optimized for critical coupling ($\beta=1$) by a slight rotation, i.e. modifying its effective surface area to the H-field in the cavity.

\begin{figure}
\centerline{\psfig{file=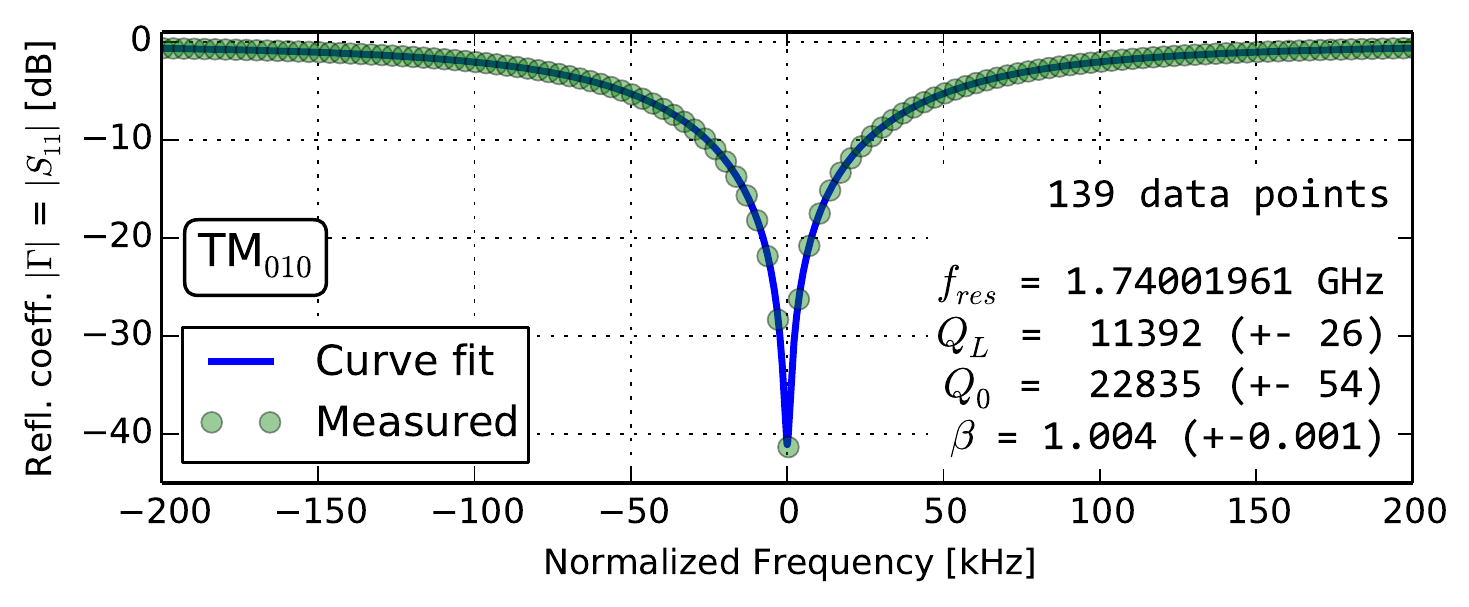, width=1\linewidth}}
\caption{
Result of a $S_{11}$ measurement of the detecting cavity with a VNA, immediately before the ALPs run. The cavity parameters have been extracted by means of a curve fitting procedure, adapted from \cite{src:cavQfitting}. 
}
\label{fig:cavFit}
\end{figure}

The loaded quality factors significantly influence the sensitivity of the experiment and therefore had to be determined with high accuracy. The frequency dependent reflection coefficient $\Gamma$ has been measured with a Vector Network Analyzer (VNA). The cavity parameters $Q_0$, $Q_L$, $\beta$ and their respective uncertainties have been extracted from the VNA data by means of a curve fitting algorithm, described in \cite{src:cavQfitting}. The resulting quality factors for the HSP and ALP measurement runs can be found in Table~\ref{tbl:paramHSP} and Table~\ref{tbl:paramALP}. A typical result, taken imediately before the ALPs run in June 2013 is shown in Fig.~\ref{fig:cavFit}.

The coupling loops in both cavities are made from copper wire, hence a small amount of signal power is lost due to its finite conductivity.
The coupling loss can be estimated from the reflection coefficient far-off resonance, which would be equal to $|\Gamma| = 0$~dB (a short or open circuit) in the ideal case. We measured a coupling loss in the order of $|\Gamma| = -0.1$~dB. Due to its small value and due to the fact that the coupling losses are implicitly included in the loaded quality factor ($Q_L$) for the critically coupled cavity, we do not have to consider them separately for the detection sensitivity of the experiment.

Note that both cavities provide $|\Gamma| < -30$~dB return loss on their resonant frequencies.
Hence the reflected RF-power is small and signal attenuation due to impedance mismatch at the cavity couplers can be neglected. 

\section{\label{sec:resMon}Monitoring of resonant frequency}
Maximum sensitivity to a WISP related signal can only be achieved if the resonant frequency ($f_{\mathrm{res}}$) of both cavities agrees with the system frequency ($f_{\mathrm{sys}}$), which is the frequency of the emitting cavity drive signal.
The cavities are sensitive to temperature variations due to the thermal expansion and contraction of their wall materials.
Sensitivity will degrade if a cavity drifts off-frequency during the measurement run.
This could disguise a potential WISP signal and lead to an invalid exclusion result. Therefore it is critical to monitor the instantaneous resonant frequency of both cavities, take the maximum deviation into account and estimate a worst case degradation in detection sensitivity.

The resonant frequencies have been monitored by recording three different observables during the measurement runs:
\begin{description}
\item[RF power] The emitting cavity has been monitored by logging the incident ($P_{\mathrm{inc}}$) and reflected RF power ($P_{\mathrm{refl}}$) at the cavities coupling port. Reflected power will only be minimum if $f_{\mathrm{sys}} = f_{\mathrm{res}}$. Detuning causes an increase in $P_{\mathrm{refl}}$. If completely off-tune, all of the incident RF power would be reflected \cite{src:CAS}.
$P_{\mathrm{inc}}$ and $P_{\mathrm{refl}}$ are measured on a directional coupler, placed on the coaxial line between power amplifier and cavity.
The coupled signals are converted to DC voltages proportional to RF power by detector diodes and recorded with a data logging device (Picolog ADC-20). A VNA was used to calibrate the setup, allowing absolute power levels to be recorded.
\item[Noise power] For the detecting cavity, we evaluate the spectral noise power density $N_{\mathrm{o}}$ around $f_{\mathrm{res}}$.
As the noise temperature of the cavity walls (298~K) is significantly higher than the noise temperature of the amplifier (43~K), a good estimate of the absolute resonant frequency can be determined from the maximum of the noise power spectrum.
A dedicated spectral noise measurement with a span of 1~MHz has been carried out with the VSA before and after each experimental run.
Note that the VSA is already connected to the receiving cavity for the purpose of recording experimental data and no changes to the hardware were necessary for this measurement.
Furthermore, the data during the measurement run has been evaluated for its average noise power over time. The trace shows a maximum if the cavity is on resonance.
\item[Physical temperature] The physical temperature of both cavities has been measured with high precision by two LM35 sensors. The change in resonant frequency is directly proportional to the change in temperature.
We have measured the proportionality constant beforehand (see Table~\ref{tbl:tempConst}), which allows us to make a statement about the maximum deviation of the resonant frequency during the measurement run.
\end{description}

\begin{table}
\centering
\caption{Measured $\Delta f / \Delta T$ of two cavity modes.}
\label{tbl:tempConst}
\begin{tabular}{c c}
\toprule
TM$_{010}$:~-33.5~kHz/$^\circ$C \quad & \quad TE$_{011}$:~-57.2~kHz/$^\circ$C\\ 
\toprule
\end{tabular}
\end{table}

\section{\label{sec:cavStab}Cavity operation}
The emitting cavity dissipates up to 50~W of heat by forced air cooling without any external temperature stabilization. Before data taking, the cavity was heated by RF power, while the tuning screw was continuously adjusted to keep it on resonance. After approximately 1~h, the cavity reaches thermal equilibrium and no further tuning is necessary.
Once in this state, no major frequency drift occurs because of a feedback process:
An increase in cavity temperature manifest itself in expansion -- hence in a lower resonant frequency, which in turn leads to less RF power being absorbed by the cavity; the temperature of the cavity decreases, resulting in a stable feedback operation. This stability can only be achieved on the upper half of the resonance curve. To stay within that region, even with small fluctuations of ambient temperature,  $f_{\mathrm{sys}}$ has been set slightly higher than $f_{\mathrm{res}}$, leading to $\approx 3~$W of constantly reflected RF power. The actual absorbed RF power in the cavity, taking reflection losses into account, is $P_{\mathrm{em}} = P_{\mathrm{inc}} - P_{\mathrm{refl}}$. The average of $P_{\mathrm{em}}$ during the measurement run has been utilized for the exclusion limit calculation.

\begin{figure}
\centerline{\psfig{file=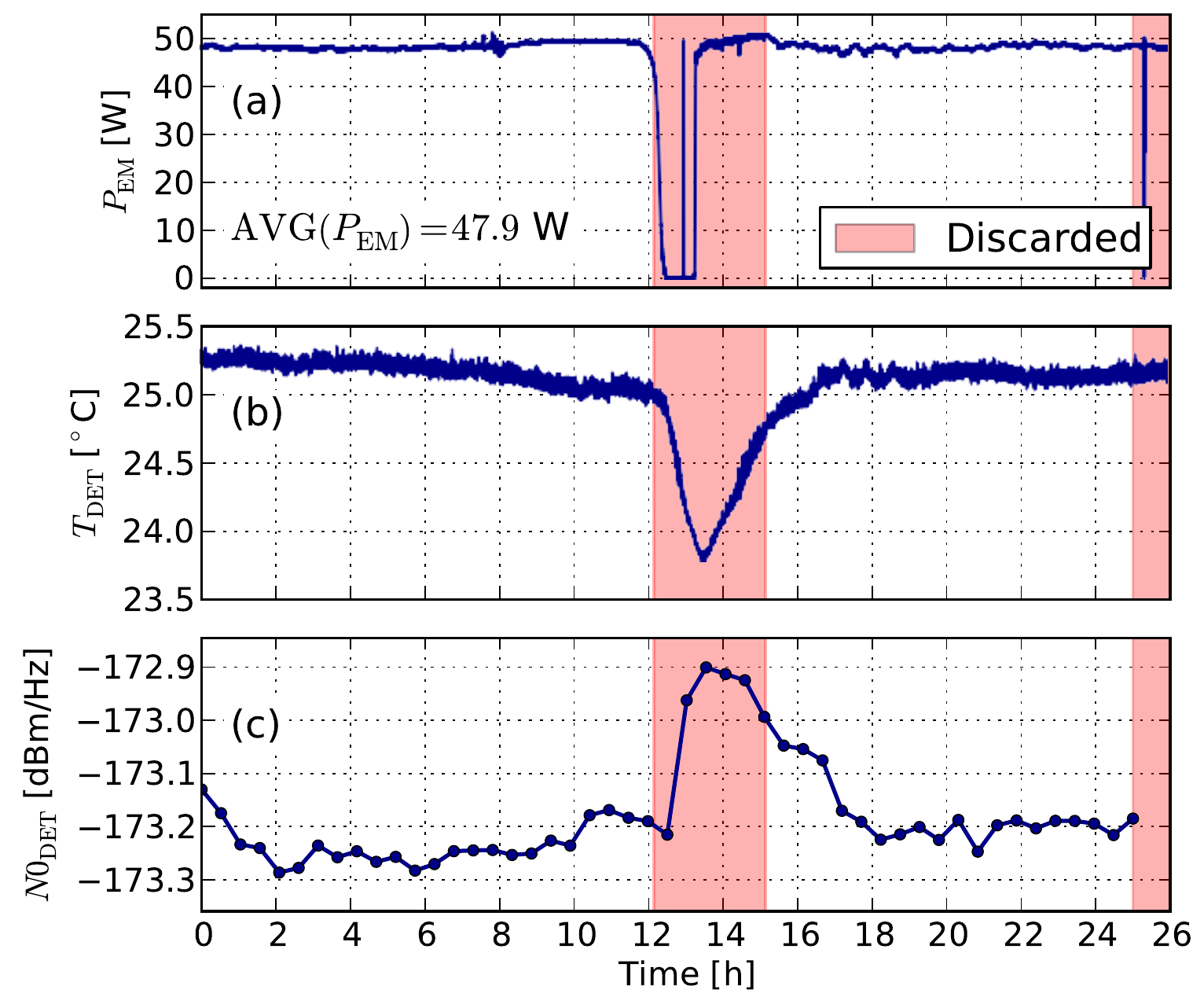, width=1\linewidth}}
\caption{(a): emitting cavity absorbed RF power (defined as $P_{\mathrm{EM}} = P_{\mathrm{inc}} - P_{\mathrm{refl}}$), (b): Physical temperature of the detecting cavity, (c): Noise power density from the detecting cavity. Data was recorded during the ALPs run in June 2013.}
\label{fig:tempAndRF}
\end{figure}

Figure~\ref{fig:tempAndRF}(a) shows a trace of the measured $P_{\mathrm{em}}$ for the ALPs run in June 2013.
During this run, an unexpectedly large fluctuation of the ambient temperature resulted in a thermal runaway condition after the first 12~h of data taking. The emitting cavity drifted off-resonance, reflected all incident RF power and cooled down to ambient temperature within a few minutes.\\
After noticing this condition, it was possible to bring the cavity back to the nominal operating temperature and resonant frequency by adapting $f_{\mathrm{sys}}$ remotely. We were able to continue the experiment after a 3~h period, during which the recorded data had to be discarded. Despite the thermal runaway incident, this run still yields highest sensitivity towards ALPs.

\begin{figure}
\centerline{\psfig{file=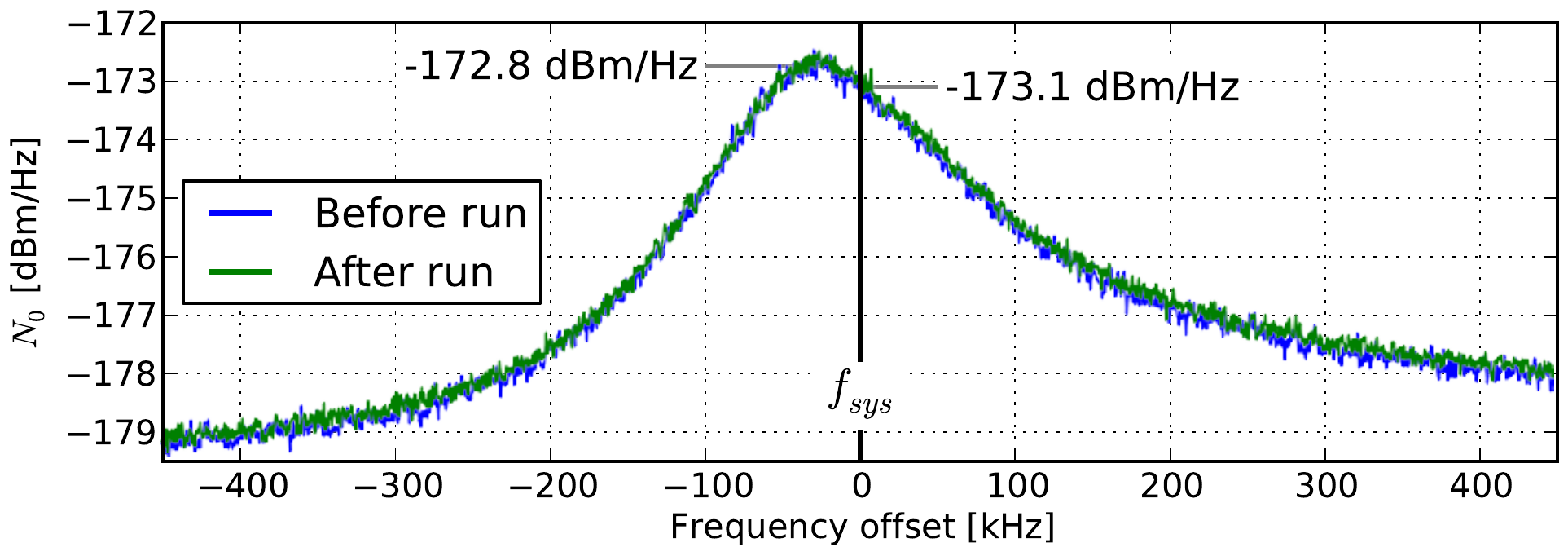, width=1\linewidth}}
\caption{
Noise power density $N_0$ at the coupling port of the detecting cavity, indicating its resonant frequency in relation to $f_{\mathrm{sys}}$. The measurement has been done immediately before and after the 25 h ALPs run in June 2013.}
\label{fig:detCavTune}
\end{figure}

The detecting cavity was only tuned at the beginning of the measurement run. After closing its shielding enclosure, the tuning screw is not reachable and was left untouched. A good indication of its resonant frequency at the beginning (t=0~h) and end (t=25~h) of the measurement run is given by the maximum of its spectral noise power density ($N_0$) in Fig.~\ref{fig:detCavTune}. The cavities $f_{\mathrm{res}}$ is $\approx 20$~kHz below $f_{\mathrm{sys}}$. The noise power at $f_{\mathrm{sys}}$ is $\approx 0.3$~dB below the maximum. We would expect the same amount of degradation for a hypothetical WISP signal. There is no visible difference between the state of the cavity at the beginning and end of the measurement run. However, the temperature measurement, shown in Fig.~\ref{fig:tempAndRF}(b) indicates a significant change of $\Delta T = -1.25 ^\circ$C at t=12~h. The cause for this fluctuation was the unexpected cool-down of the emitting cavity, which was in close vicinity to the detecting cavity.
According to Table~\ref{tbl:tempConst}, the change in temperature corresponds to a relative change in resonant frequency of $\approx \Delta f = + 40$~kHz, which would place the cavities resonance $\approx 20$~kHz above $f_{\mathrm{sys}}$ and cause a worst case reduction of -0.3~dB in signal power. 
Note that it was not practical to measure the actual temperature of the detecting cavity, as its EMI shielding enclosure would have been compromised by the copper wires of the temperature sensor. Instead, the sensor has been placed on the outside wall of the shielding enclosure, which is in good thermal contact with the cavity. The actual fluctuations of cavity temperature are therefore less than the measured $T_{\mathrm{DET}}$.

As a further crosscheck, the average noise power in a bandwidth of 2~kHz around $f_{\mathrm{sys}}$ has been evaluated from the recorded experimental data. The result is shown in Fig.~\ref{fig:tempAndRF}(c). The noise power density at t=0~h is $N_0 = -173.1$~dBm/Hz, which is in good agreement with the blue trace in Fig.~\ref{fig:detCavTune}. At t=12~h, an excursion of +0.3~dB is visible. This indicates a shift of the resonance curve by $\approx \Delta f = + 20$~kHz, centering it on $f_{\mathrm{sys}}$. At t=18~h, the cavity has warmed up again and reached its original and slightly detuned state, which is in good agreement with the green trace in Fig.~\ref{fig:detCavTune}.\\
In conclusion, we can make the following statements for the ALPs run in June 2013:
\begin{itemize}
  \item The worst case signal degradation of an hypothetical ALP signal due to detuning of the detecting cavity was $\leq 0.3$~dB~$= 7\%$.
  \item The average RF power absorbed by the emitting cavity was $P_{\mathrm{em}} = 47.9$~W
\end{itemize}


The same monitoring principles have been applied during the HSP run in September 2013. During this run no thermal runaway condition occurred, and the emitting cavity was stable throughout the entire recording time, lasting 3 x 29 h.

\section{\label{sec:emShield} Electromagnetic shielding}
Shielding is critical around the detecting cavity and the microwave receiver to eliminate electromagnetic interference (EMI) from ambient sources like cellphones or wireless network transceivers. 
Shielding is also necessary to avoid coupling between the two cavities by electromagnetic (EM) leakage, which has to be attenuated below the detection threshold of the microwave receiver. Leaking photons would generate false results, as this kind of signal could not be distinguished from a signal propagating by WISP conversion.
From the expected EM field strengths in emitting (180~kV/m) and detecting cavity (20~nV/m), we can get an estimate for the required amount of shielding. At only 20~mm separation between the two cavities, the fields need to be attenuated by $> 10^{13} = 260 \mathrm{~dB}$ to ensure meaningful results. Most microwave components used in the setup like SMA connectors or semi-rigid coaxial cables provide less than 120 dB of shielding, making an external shielding enclosure and strategic use of optical fibres for signal transmission necessary.

\begin{figure}
\centerline{\psfig{file=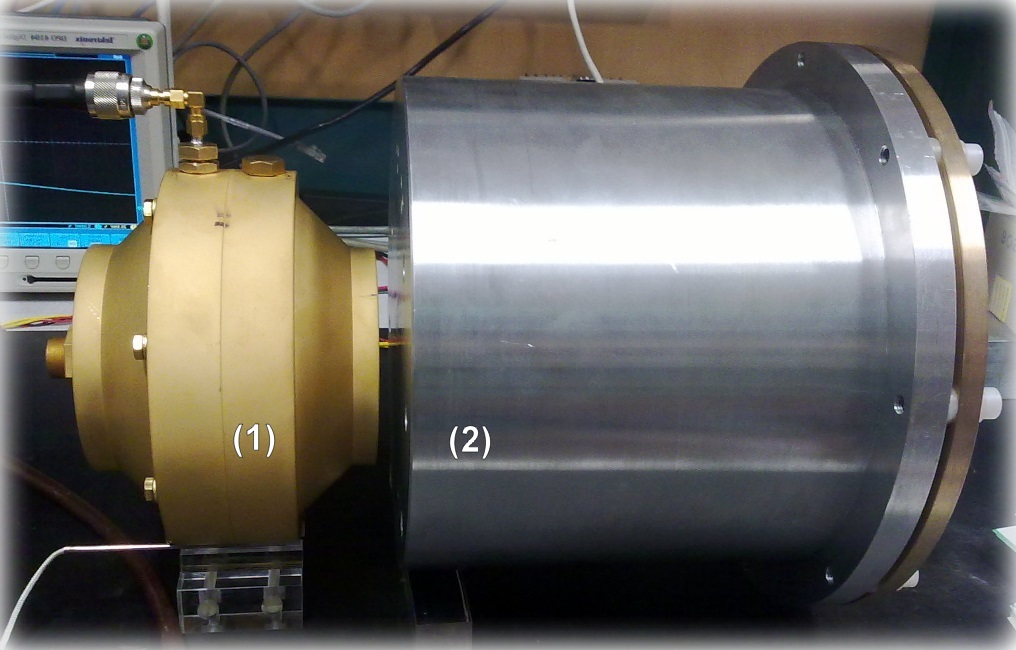, width=1\linewidth}}
\caption{
Photo of emitting cavity (1) and shielding enclosure (2) containing the identical detecting cavity. For ALP search, both parts were placed in the bore of a solenoid magnet with the same arrangement as shown in the picture.
}
\label{fig:magnParts}
\end{figure}

The EM shielding has been split into two separate enclosures. One is placed in the magnet, housing the detecting cavity and the RF frontend, as shown on the right hand side of Fig.~\ref{fig:magnParts}. The second shielding enclosure is placed outside of the magnet and contains the instrumentation needed to detect the weak microwave signals. Both enclosures have been lined with microwave absorbing foam on their inside walls. This dampens resonances, which could lead to a degraded shielding performance at certain frequencies \cite{src:EMIcavRes}. The RF signals between the two shielding boxes are transmitted by optical fibres utilizing analog transceivers. An optical ethernet link is used for remote controlling the signal analyser. Optical fibres have two distinct advantages in this application:
\begin{itemize}
  \item Compared to coaxial transmission lines, they provide galvanic isolation and a nearly infinite shielding attenuation. Microwave interference does not influence the optical carrier and can not couple into the shielded domain.
  \item Optical fibres are free of metals, making them efficient with the tubular waveguide style feedthroughs \cite{src:EMVgrundlagen} used in both shielding enclosures.
\end{itemize}

\begin{figure}
\centerline{\psfig{file=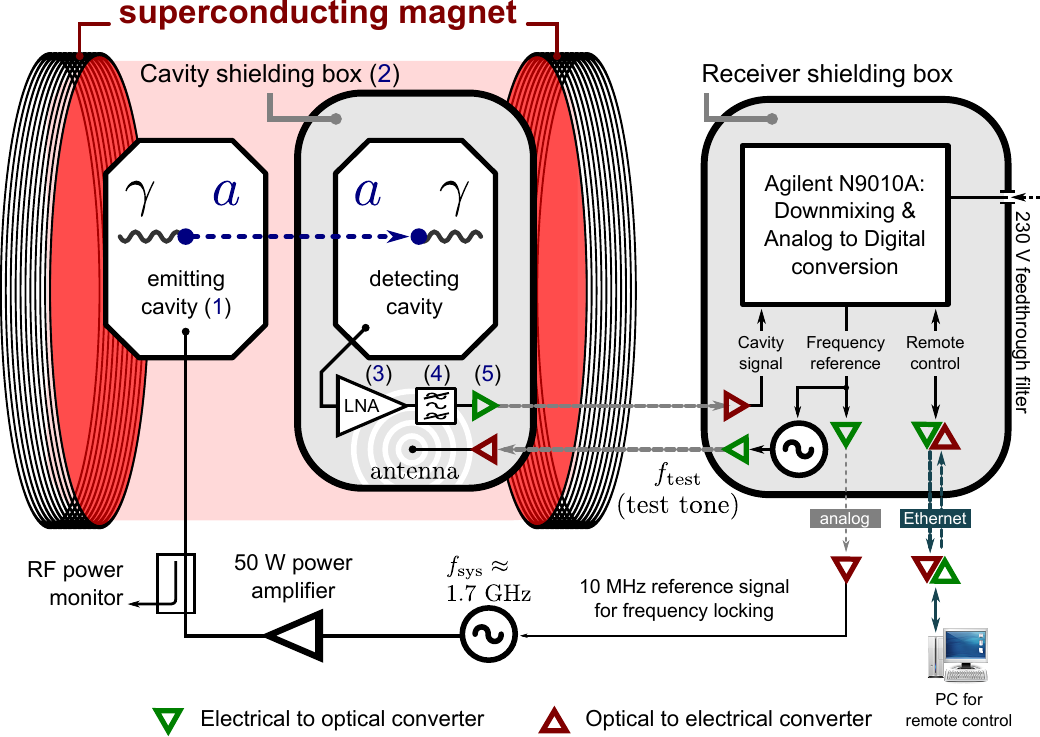, width=1\linewidth}}
\caption{Detailed block diagram of the experimental setup.}
\label{fig:OverView}
\end{figure}

A detailed schematic of the experimental setup is shown in Fig.~\ref{fig:OverView}.
To measure the shielding effectiveness, a microwave source has been placed within the enclosure under test and the electric field strength at several fixed points outside the enclosure has been compared with its lid open and closed \cite{src:EMVgrundlagen}. The field strength was measured with a calibrated electric field probe, which also makes it possible to quickly localize weak spots in the shielding. Both enclosures provide $\approx 90$~dB and each of the cavities provide an additional $\approx$~110 dB of shielding. Therefore the combined EM attenuation of the experimental setup is $\approx 310$~dB, making thermal noise the limiting factor for the minimum detectable signal.

For diagnostic purposes, a sinusoidal signal of known frequency is emitted within the shielding enclosure during each measurement run. This ``test tone'' of relatively low and constant power ($\approx -100$~dBm) couples from a $\lambda / 4$ antenna to the detecting cavity and to the components of the receiver frontend. By identifying the signal in the recorded spectrum, we demonstrate that the entire signal processing chain was operational during the measurement. This also allows to evaluate unwanted frequency offsets, frequency drifts, or phase noise by comparing shape and position of the measured signal peak to the expected one. 
Furthermore, the observed power of the test tone is used as an indicator for a major degradation in the EM shielding.
For example, a faulty RF connector on the detecting cavity can lead to excessive RF leakage. The observed test tone power would increase by several orders of magnitude, which is an immediate indication for a fault condition. 
The test tone was transmitted over an optical fibre into the shielding enclosure, using a reverse biased photo diode (Hamamatsu G9801) to convert the optical to an electrical signal. The test tone frequency $f_{\mathrm{test}}$ has been offset by $\approx 400$~Hz to $f_{\mathrm{sys}}$, which avoids any interference with WISP detection.

\section{\label{sec:front} RF frontend}
All components of the RF frontend are mounted within the cavity shielding enclosure and need to be compatible with the 3~T magnetic field. The noise like signal from the detection cavity is amplified, filtered, modulated on an optical carrier and transmitted over an optical fibre to the shielding box outside the magnet. 
\begin{figure}
\centerline{\psfig{file=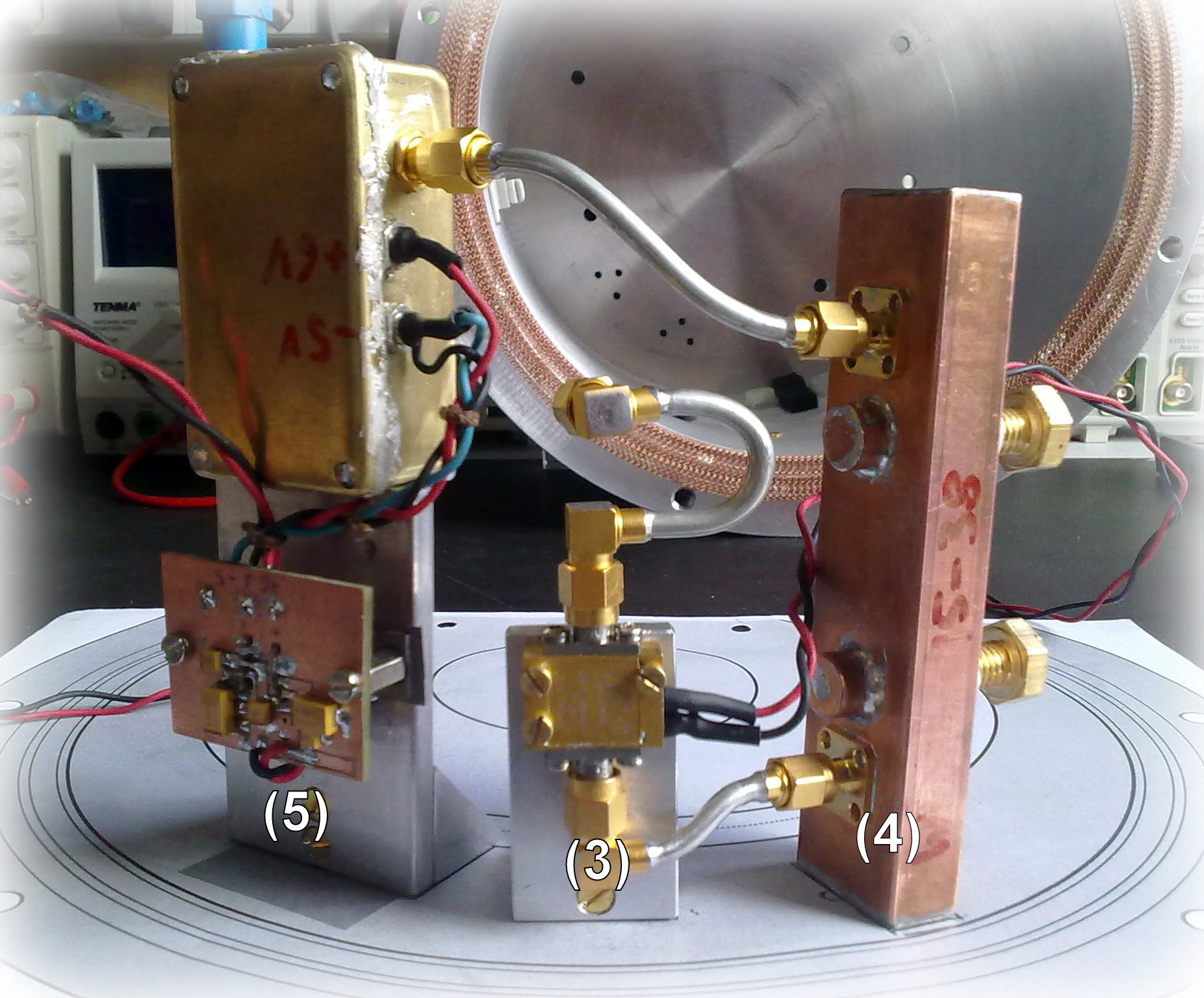, width=1\linewidth}}
\caption{The components of the RF frontend, unmounted from the cavity shielding box. From left to right: Analog optical link, low noise amplifier and bandpass filter.}
\label{fig:noiseRed}
\end{figure}

The low noise amplifier (LNA) of type MITEQ AMF-3F-0200-400-06-10P has been tested and characterized in the 3~T magnet. This was necessary, as some components in the amplifier might be affected by a high level magnetic field \cite{src:LNA_Magnet}. The LNA provides a nominal gain (G) of 45 dB and a noise figure (NF) of 0.6 dB at $f_{\mathrm{sys}}$. Its equivalent noise temperature is $T_{\mathrm{LNA}} = 43$~K, making thermal noise from the detecting cavity the only significant noise source in the receiving chain.  

The preliminary HSP measurement runs until December 2012 have been successfully completed using a commercial optical transmitter, type MITEQ LBT-50K4P5G-25-15-M14. However the module received permanent damage after first tests in the 3 T magnet in preparation for the run in June 2013. As no magnetically compatible replacement product was readily available from industry, a commercial satellite TV low noise block from the company INVACOM was adapted for our purpose. Ferromagnetic materials of significant mass, including all ferrite cored inductors have been removed, the internal DC/DC converter has been replaced with a magnetically compatible power supply. The optical link was measured and achieved a nominal noise figure of 10 dB and a gain of 19 dB in the frequency range of 0.5 GHz - 4 GHz, which are of comparable performance to the MITEQ link. It proved to operate reliably in a magnetic field of up to 3 T.

It is necessary to calibrate the measured power spectra to obtain the absolute noise power at the coupling port of the detecting cavity. The so called ``hardware transfer function'' (HTF) had to be determined. It corresponds to the cascaded gain and noise figure of all frontend components and cables between cavity and signal analyzer. To minimize errors due to thermal drifts and influences of the magnetic field, the HTF has been measured in the magnet, immediately before the WISP measurement run.
The HTF was determined by the Y-factor method \cite{src:AgilentNoise}. For this purpose, the detecting cavity was disconnected from the LNA and a calibrated noise diode was connected through a 5 m long cable. The cable was necessary to prevent interference of the noise diode due to the magnetic field. The exact attenuation of the cable has been determined beforehand and taken into account. The HTF of the receiving chain has been determined as NF = $0.7 \pm 0.2$ dB and G = $60.4 \pm 0.5$ dB at $f_{\mathrm{sys}}$. The measurement uncertainty has been estimated by a method described in \cite{src:AgiNFuncPaper}. 

During earlier HTF measurements, a saturation of the optical transmitter was observed due to the output of the 15 dB ENR noise diode ($T_N = 2300$~K) being amplified by the LNA over a wide bandwidth $>2$ GHz. A non-magnetic, adjustable bandpass filter was designed, built and inserted between LNA and optical transmitter. The filter substantially reduces noise-power by limiting the frontend bandwidth to 25~MHz, preventing this saturation and effectively increasing the dynamic range. The filter is based on an evanescent mode design described in \cite{src:EvanescentFilter,src:WGFilters}.

\section{\label{sec:proc} Data processing and evaluation}
The signal from the RF frontend is recorded by an Agilent EXA N9010A signal analyzer. The center frequency was set to $f_{\mathrm{sys}} + 400~\mathrm{Hz}$ to avoid internal spurious signals appearing at the important parts of the spectrum \cite{src:moiIPAC12}. The instrument shifts the center frequency to baseband before digitizing and recording the complex quadrature signal with a bandwidth of 2 kHz.

For offline data processing, the spectral power of the recorded noise like signal is estimated by a python script.
For the ALPs run in June 2013, the time record had to be divided into two 10 h long continuous segments, discarding a 3~h long segment of data where the emitting cavity drifted off tune.

\begin{figure*}
\centerline{\psfig{file=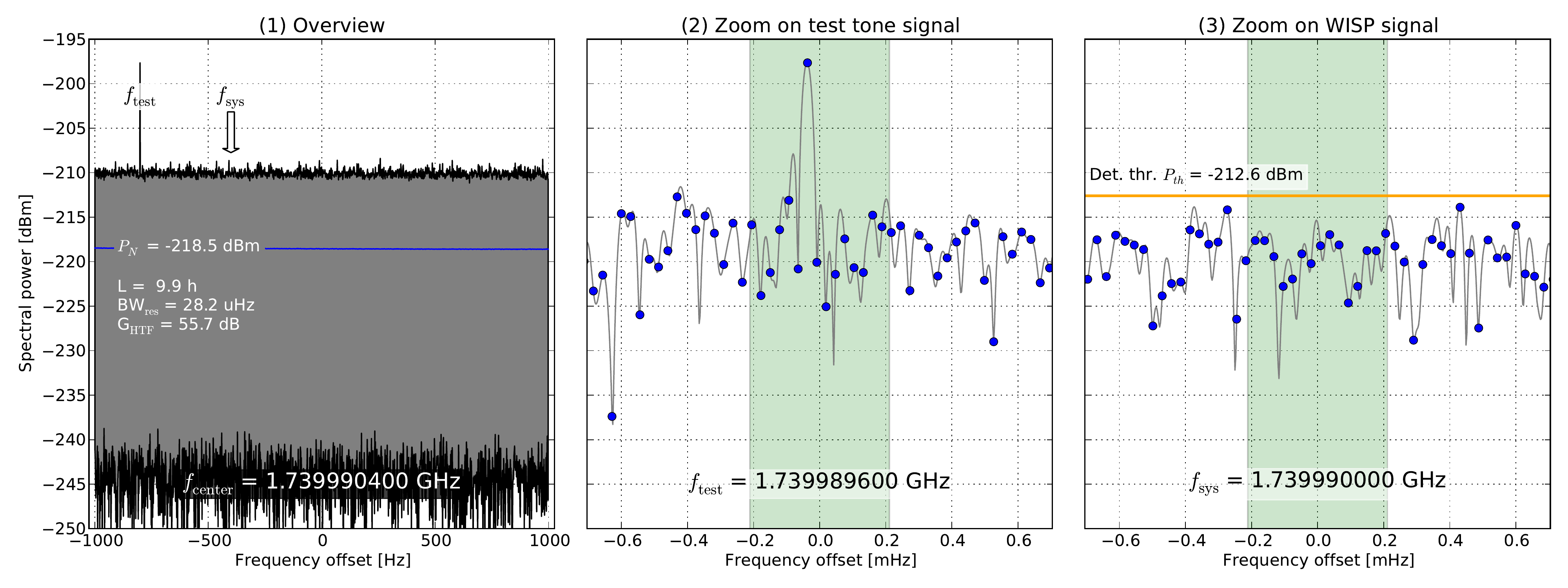, width=1\linewidth}}
\caption{Resulting power spectrum from the ALP run in June 2013. (1) Overview over full recorded span. (2) Zoom on $f_{\mathrm{test}}$, where the test tone signal is clearly visible. (3) Zoom on $f_{\mathrm{sys}}$, where no WISP signal is visible. The green shaded areas mark the frequency range where a signal would be expected. Blue dots indicate the spectral bins calculated by a direct DFT operation, the grey lines show the underlying continuous spectrum, approximated by zero padding the time domain data. All three plots share the same Y - axis}
\label{fig:resultSpect}
\end{figure*}

The complex spectra of each segment are calculated by an Discrete Fourier Transform (DFT), efficiently implemented by the software library FFTW \cite{src:FFTW05}. The two subspectra are averaged, resulting in the final spectrum in Fig.~\ref{fig:resultSpect} (1). It consists of $\approx 71 \cdot 10^6$ spectral bins, which have been decimated to 1500 points in the overview plot, showing minima, maxima (as grey area) and average values (as a blue line) of each group of spectral bins. This substantially reduces the amount of data handling while preserving sharp peaks or sudden excursions, which are the expected signatures of a WISP signal.

The DFT represents a matched filter for detecting sinusoidal signals in white background noise \cite{src:statPaper2} and is therefore the most efficient algorithm for this purpose. Each spectral bin resulting from the DFT operation can be seen as the integrated output power of a bandpass filter, tuned to a specific frequency. 
The bandwidth of each filter is determined by BW$_{\mathrm{res}} = 1/l$, where $l$ is the length of the recorded time trace.
The displayed average background noise level in each spectral bin is given by $P_N = \mathrm{BW_{res}} N_0$, where $N_0$ is the noise power density of the input signal. 
A pure sinusoidal signal at fixed frequency (like the one we would expect from a WISP) has an infinitely narrow bandwidth and will always deposit its entire signal power ($P_{\mathrm{sig}}$) within one single spectral bin. Therefore the signal to noise ratio in this bin, defined as S/N = $P_{\mathrm{sig}}/P_N$, is proportional to the length of the recorded time trace. This is in contrast to averaging $n$ spectra, where S/N only improves by a factor of $\sqrt{n}$.

No window function was used before calculating the FFT. This yields the most narrow resolution bandwidth BW$_{\mathrm{res}}$, the lowest $P_N$ and the largest possible S/N for detecting sinusoidal signals \cite{src:fftWind,src:FFTpaper}. Note that window functions are often used to diminish the effects of spectral leakage and provide a steep fall-off around signal peaks. This is not required in our case as we do not need to resolve signals in the spectrum which are tightly spaced in frequency or which have a high dynamic range.

\begin{figure}
\centerline{\psfig{file=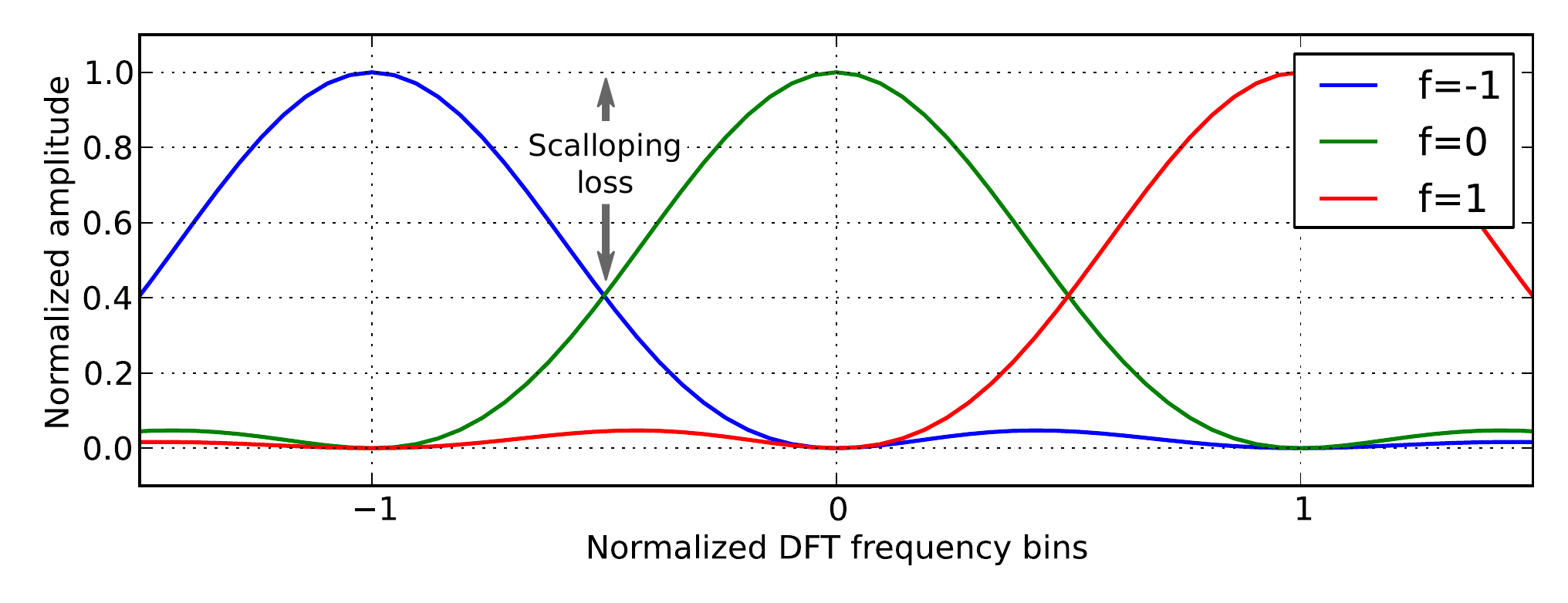, width=1\linewidth}}
\caption{Response of three spectral bins without zero padding. The amplitude of a signal falling between the bins can be attenuated by a factor of 0.64 (-3.92~dB).}
\label{fig:scallop}
\end{figure}

The resulting spectral estimate suffers from certain artifacts, originating from the definition of the DFT. In our case, the most critical one is the so called scalloping loss, which can be observed if a sinusoidal signal falls between two frequency bins in the spectrum. Its amplitude can be attenuated by up to 3.92 dB \cite{src:FFTpaper}. This is illustrated by Fig.~\ref{fig:scallop}. 
A signal, sampled in the time domain, corresponds to a continuous spectrum in the frequency domain. Scalloping loss occurs because the DFT defines the minimum number of sampling points on this underlying continuous spectrum, without causing information loss (Nyquist rate in the frequency domain). One way to avoid scalloping loss is to use a flat-top window. However, this trades resolution bandwidth for amplitude accuracy, which would reduce the detection sensitivity.\\
A better way is to calculate more than the minimum number of frequency bins. This can be achieved by zero padding the time domain signal before the DFT operation. The underlying continuous spectrum is sampled more frequently in the frequency domain, which leads to a more accurate representation of a signal peak, if it falls between two frequency points. For the data analysis, zero padding has been applied with 10 times the number of measured samples, reducing scalloping losses to a negligible amount. For the two zoomed spectra in Fig.~\ref{fig:resultSpect} (2) and (3), the interpolated data is shown as a grey line, while the sampling points from a DFT operation without zero padding are shown as blue dots.

As we effectively define a very narrowband filter around the WISP signal, the long-term frequency stability of the RF-source, the signal analyzer and any other oscillator involved in the receiving chain can critically influence the detection sensitivity.
It has been demonstrated in \cite{src:narrowband} that excessive frequency drifts would smear out the sharp peak we would expect from a sinusoidal signal in the spectrum. The power of a hypothetical WISP signal would spread over several frequency bins and the signal to noise ratio would degrade.
To ensure the WISP signal stays within one frequency bin of width $BW_{\mathrm{res}}$, we require a fractional frequency accuracy of $\Delta f / f  = BW_{\mathrm{res}} / f_{\mathrm{sys}} \approx 2 \cdot 10^{-14}$ during the whole measurement time. 
Note that this requirement only applies to the frequency accuracy of the oscillators relative to each other. Frequency drifts which apply to all oscillators in the same way will cancel out in the result.
We have achieved the required stability by phase-locking all oscillator to a common 10~MHz frequency reference. 
Long term phase drift measurements showed, that the synchronized RF sources are precise enough for the narrowband measurement and no broadening of the signal peak is expected. This has been confirmed by the power spectra of each experimental run, where a test tone signal is visible as narrow peak with the minimum possible width dictated by the DFT.

\begin{figure}
\centerline{\psfig{file=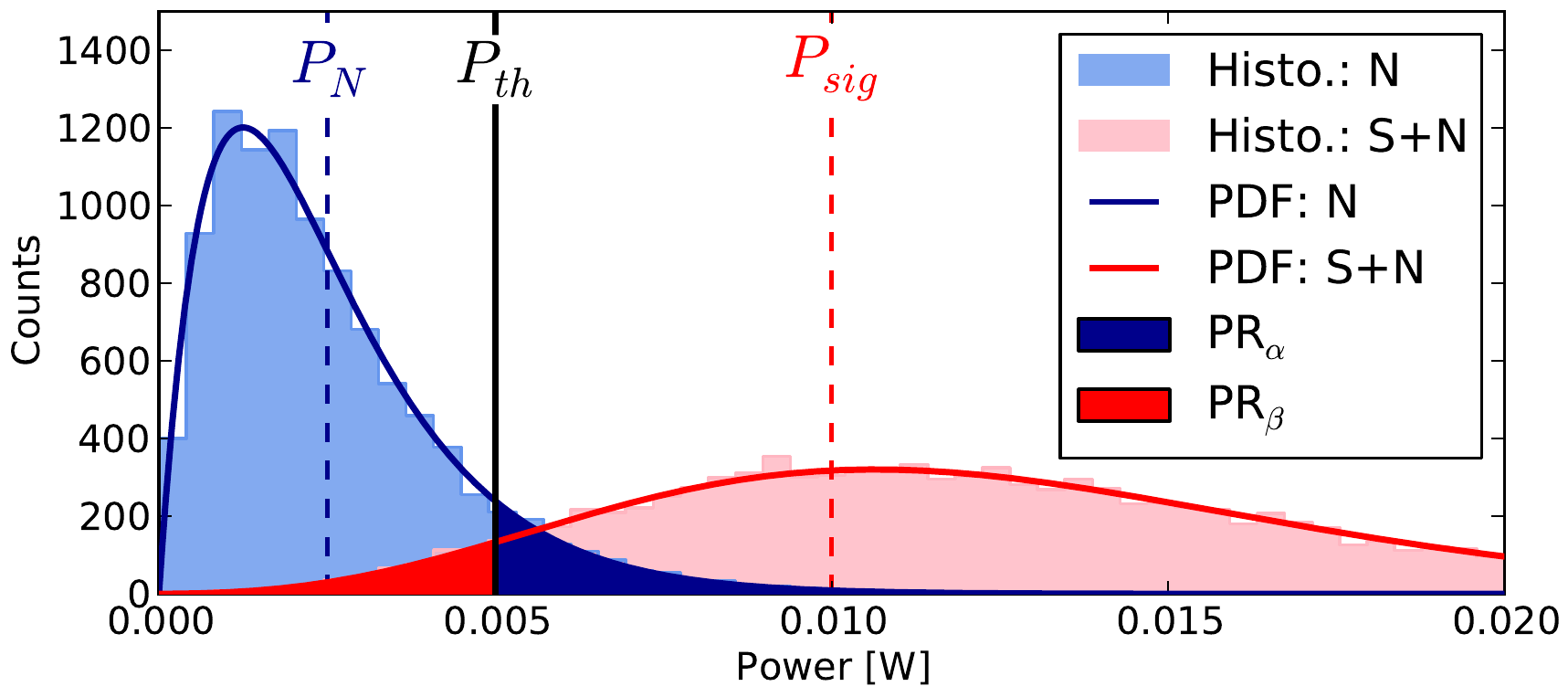, width=1\linewidth}}
\caption{Monte Carlo simulation of the fluctuation of one spectral bin governed by background noise (N) and another bin with an additional sinusoidal signal (S+N). The analytical PDFs show good agreement with the histogram of the Monte Carlo data. Furthermore, an exemplary detection threshold $P_{th}$ and the resulting error probabilities $PR_{\alpha}$ and $PR_{\beta}$ have been indicated. Note that the power levels and probabilities are arbitrary and do not correspond to a measurement run.}
\label{fig:monte}
\end{figure}

The resulting power spectrum from the ALPs run in June 2013 is shown in Fig.~\ref{fig:resultSpect}.
The test tone signal is visible as a single peak, spanning only one single bin. Its absolute position on the frequency axis was offset by $\approx 34~\mu$Hz due to the finite frequency resolution of the RF oscillators. To accommodate for these offsets, a window has been defined with a width of $15 \cdot \mathrm{BW_{res}} = 423~\mu$Hz around the frequency where an ALP signal would be expected.

To decide the outcome of the experiment, we compare the power level of each frequency bin within this window to a predefined detection threshold $P_{\mathrm{th}}$. Defining the threshold is an exercise in statistical hypothesis testing. We specify:
\begin{description}
\item[$H_0$:] Null hypothesis. There is no WISP signal with a power of $\geq P_{\mathrm{sig}}$ and the frequency bins are governed exclusively by background noise.
\item[$H_1$:] Alternative hypothesis. There is a WISP signal added to the background noise with a power of $\geq P_{\mathrm{sig}}$.
\end{description}
The probability density functions (PDF) in both cases are known. Under $H_0$, the relevant frequency bin in the power spectrum will obey a central $\chi^2$ distribution. Under $H_1$, it will obey a non - central $\chi^2$ distribution, where the non-centrality parameter equals the WISP signal power $P_{\mathrm{sig}}$.
Both distributions have two known parameters: the degrees of freedom is $k=4$, because the power spectrum is calculated from the magnitude of independent real and imaginary parts, which have been averaged twice. Furthermore, the parameter $\sigma$, describing the variance of the distributions, is related to the average noise power by $\sigma = P_N / k$. As $P_N$ can be estimated from the frequency bins where no WISP signal is expected, this parameter is considered known.
Details to the derivation of the PDFs are given in \cite{src:statPaper1, src:statPaper2, src:statPaper3}. The analytical PDFs have been cross-checked with a Monte Carlo simulation. We computed 10000 power spectra of synthetic input data and computed the histogram of two specific frequency bins -- one governed by noise and one with an additional signal. The corresponding analytical PDFs agree well with the histograms, as shown in Fig.~\ref{fig:monte}.

We define the following probabilities for the hypothesis test:
\begin{description}
\item[PR$_\alpha' := 5\%$] is the probability for a false positive outcome of the experiment under the assumption of $H_0$ (we have  discovered the WISP mistakenly). Note that we have to take the size of the WISP window into account. To obtain a false detection probability of 5~\% with respect to testing all 15 frequency bins, we need to set PR$_\alpha = $PR$_\alpha' / 15$.
\item[PR$_\beta := 5 \%$] is the probability for a false negative outcome of the experiment under the assumption of $H_1$ (we have excluded the WISP mistakenly).
\end{description}
The two probabilities correspond to areas under the two PDFs, above and below the detection threshold, as illustrated by Fig.~\ref{fig:monte}.
With the above definitions and the known parameters of the PDFs, we can solve the system of equations numerically and get a value for the detection threshold $P_{th}$, which corresponds to the two error probabilities.
For the ALPs run in June 2012 the detection threshold is $P_{th} = -212.6$~dBm. Figure~\ref{fig:resultSpect} shows that there is no peak exceeding this threshold within the WISP window. Therefore $H_1$ can be rejected and we can state with a confidence level of 1-PR$_\alpha' = 95$~\%, that there is no excessive signal with a power of $P_{\mathrm{sig}} \geq -210.1$~dBm in the measured data. This allows us to set an exclusion limit for ALPs.

\section{\label{sec:exclRes}Achieved exclusion results}
\begin{figure}
\centerline{\psfig{file=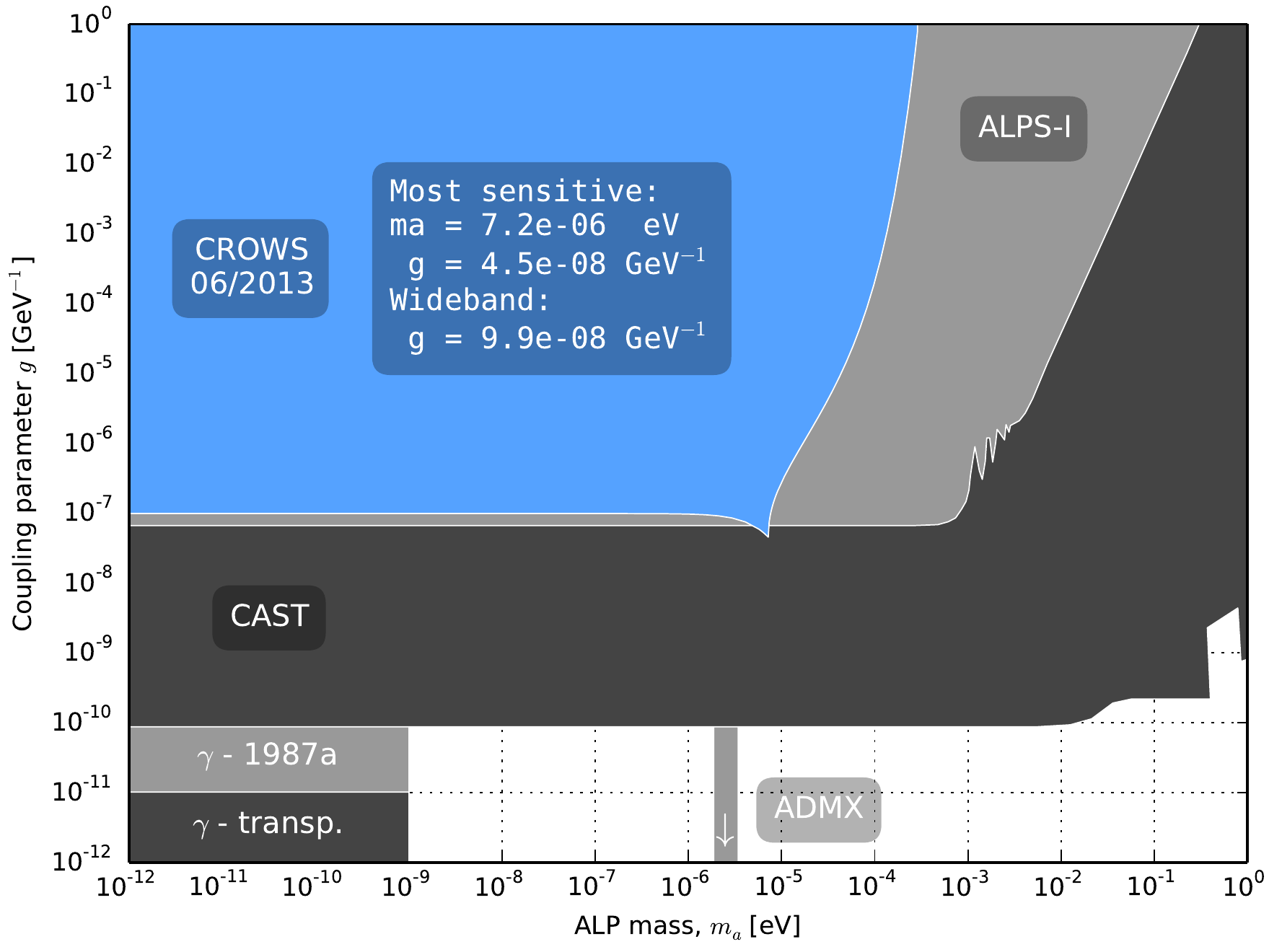, width=1\linewidth}}
\caption{CROWS: exclusion limits for ALPs for the measurement run in June 2013 in a 3 Tesla magnet. Confidence level: 95~\%.
ALPS-I: exclusion limits from the most sensitive optical LSW experiment to date \cite{src:alps1}. CAST: helioscope observing the sun to search for solar ALPs \cite{src:CAST_limits}. More details on the other experiments can be found in \cite{src:LowEnergyFrontier}.
}
\label{fig:exclPlotALP}
\end{figure}

\begin{figure}
\centerline{\psfig{file=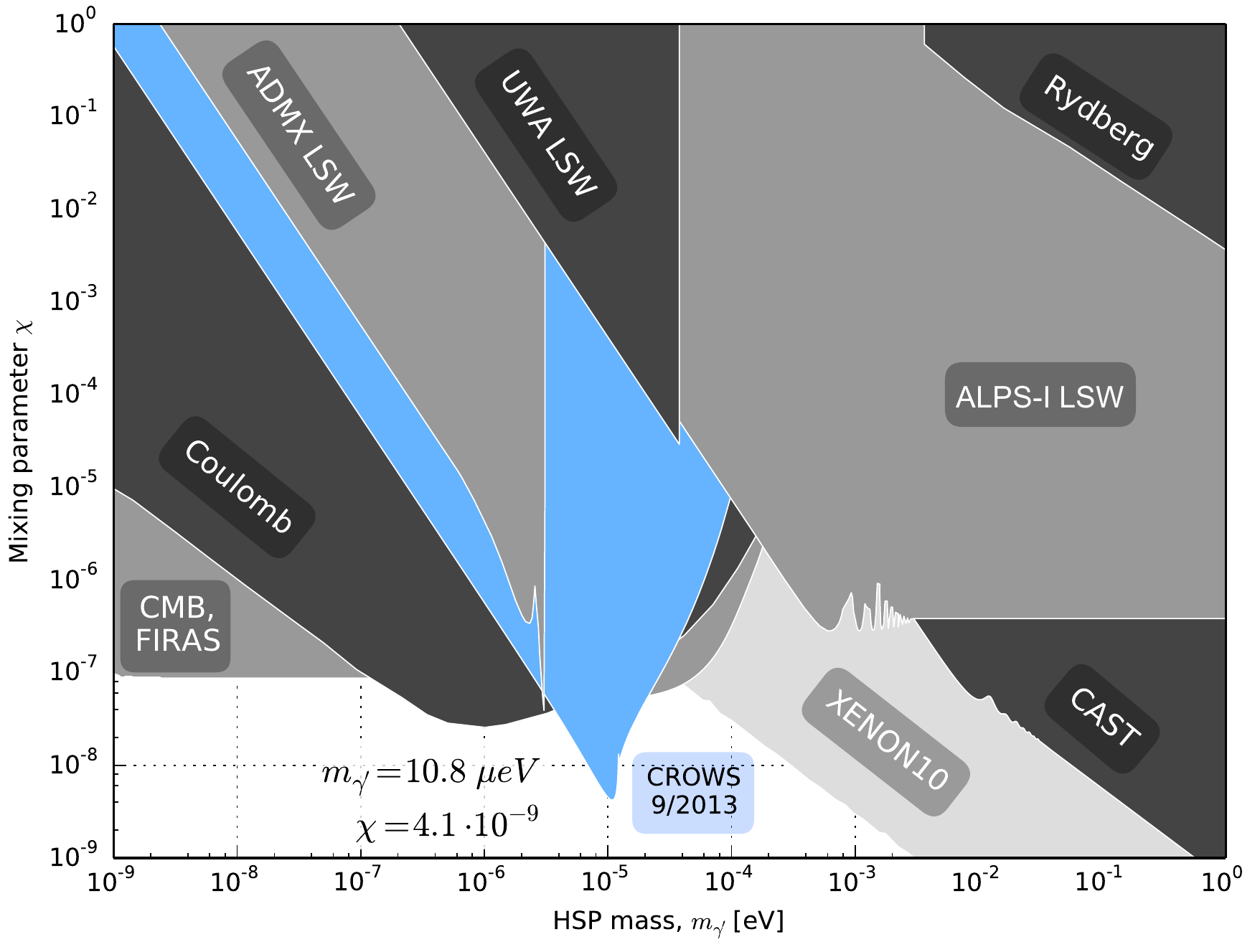, width=1\linewidth}}
\caption{CROWS: exclusion limits for HSPs for the measurement run in September 2013. Confidence level: 95~\%. UWA LSW \cite{src:UWA_LSW} and ADMX LSW \cite{src:ADMX_HSP_LSW} are similar microwave based LSW experiments.
More details on the other experiments can be found in \cite{src:LowEnergyFrontier}.}
\label{fig:exclPlotHSP}
\end{figure}

The most sensitive measurement run for HSPs has been carried out in September 2013 at CERN, recording three continuous time traces of 29~h length. For ALPs, the most sensitive run was carried out in June 2013 in cooperation with the Brain \& Behaviour Laboratory of Geneva University. We were able to operate the setup within the bore of a 3~T superconducting magnet, which is part of an MRI scanner. Over the course of one weekend, 2~x~10~h of experimental data was recorded. 
The technical parameters of these two runs have been summarized in Table~\ref{tbl:paramHSP} and Table~\ref{tbl:paramALP}.
Note that the experimental apparatus (apart from the superconducting magnet) and the method of data evaluation was identical for both measurement runs.
As no WISPs were detected, the corresponding exclusion limits in comparison to other experiments are shown in Fig.~\ref{fig:exclPlotALP} and Fig.~\ref{fig:exclPlotHSP}.

\begin{table}[t]
\centering
\caption{Parameters of the HSP run in September 2013}
\label{tbl:paramHSP}
\begin{tabular}{c}
\toprule
$f_{\mathrm{sys}} = 2.956610$~GHz \quad $Q_{\mathrm{det}} = 22739$ \quad $Q_{\mathrm{em}} = 23210$\\[0.1 cm]
$P_{\mathrm{sig}} = 3.72 \cdot 10^{-25}$ W \quad $P_{\mathrm{em}} = 35.6$ W \quad $|G|_{\mathrm{max}} = 0.51$ \\
\toprule
\end{tabular}
\end{table}
\begin{table}[t]
\centering
\caption{Parameters of the ALP run in June 2013}
\label{tbl:paramALP}
\begin{tabular}{c}
\toprule
$f_{\mathrm{sys}} = 1.739990$ GHz \quad $Q = 11392,12151$ \quad B = 2.88 T\\[0.1 cm]
$P_{\mathrm{sig}} = 9.8 \cdot 10^{-25}$ W \quad $P_{\mathrm{em}} = 47.9$ W \quad $|G|_{\mathrm{max}} = 0.94$ \\
\toprule
\end{tabular}
\end{table}

\section{\label{sec:conclusion}Conclusion}
No HSPs or ALPs were observed in the most sensitive measurement-runs of the CROWS experiment. For HSPs, the experiment was sufficiently sensitive to improve previous exclusion limits. For ALPs, it was -- in a small mass range -- more sensitive than other purely laboratory based experiments (namely laser LSW of the first generation like ALPS-1 \cite{src:alps1}) but less sensitive than extraterrestrial experiments like the CERN Axion Solar Telescope CAST \cite{src:CAST_limits}. This is the first time ALPs have been probed by a microwave based LSW experiment. Several technical challenges, like $> 300$~dB EM shielding between the cavities, keeping them frequency matched during up to 29~h long measurement runs and filtering the sinusoidal signal with a bandwidth of $BW_{\mathrm{res}} < 30~\mu$Hz to discriminate it from background noise, had to be solved. There is still significant potential for improvement as the sensitivity of the experiment scales with $B$ and $1/{f_{\mathrm{sys}}}$ for ALPs. Therefore the setup might be upgraded with a stronger magnet or lower frequency and thus larger cavities.

\section{\label{sec:Ack}Acknowledgements}
The authors would like to thank K.~Baker, P.~Blanc, A.~Collar, A.~Malagon, J.~Troschka and K.~Zioutas for a large number of inspiring discussions.
We are especially grateful to C.~Burrage for bringing the right people together at the right time, to M.~Wendt for reading the manuscript and to the BE-RF mechanical workshop at CERN for practical assistance.
Exclusion plot comparison data were provided with friendly permission from J.~Jaeckel and J.~Redondo.
We would like to express our gratitude for the support from R. Jones, E. Jensen and the BE department at CERN.

\bibliographystyle{apsrev}
\bibliography{references}

\end{document}